# Early signs of stuck pipe detection based on Crossformer

Bo Cao [a], Yu Song [a, *], Jin Yang [b], Lei Li [b]

[a] College of artificial intelligence, China University of Petroleum (Beijing), Beijing, 102249, China

[b] College of Safety and Ocean Engineering, China University of Petroleum-Beijing, Beijing, 102249, China

## Abstract

Stuck pipe incidents are one of the major challenges in drilling engineering, leading to massive time loss and additional costs. To address the limitations of insufficient long sequence modeling capability, the difficulty in accurately establishing warning threshold, and the lack of model interpretability in existing methods, we utilize Crossformer for early signs of detection indicating potential stuck events in order to provide guidance for on-site drilling engineers and prevent stuck pipe incidents. The sliding window technique is integrated into Crossformer to allow it to output and display longer outputs, the improved Crossformer model is trained using normal time series drilling data to generate predictions for various parameters at each time step. The relative reconstruction error of model is regard as the risk of stuck pipe, thereby considering data that the model can't predict as anomalies, which represent the early signs of stuck pipe incidents. The multi-step prediction capability of Crossformer and relative reconstruction error are combined to assess stuck pipe risk at each time step in advance. We partition the reconstruction error into modeling error and error due to anomalous data fluctuations, furthermore, the dynamic warning threshold and warning time for stuck pipe incidents are determined using the probability density function of reconstruction errors from normal drilling data. The results indicate that our method can effectively detect early signs of stuck pipe incidents during the drilling process. Crossformer exhibits superior modeling and predictive capabilities compared with other deep learning models. Transformer-based models with multi-step prediction capability are more suitable for stuck pipe prediction compared to the current single-step prediction models. At each time step, predictions and ground truth for each parameter, and the risk of stuck pipe are visually present to enhance the explanation of model.

# 1. Introduction

Stuck pipe is considered to be one of the most formidable challenges and costly incident in oil and gas development. Although numerous strategies have been implemented to optimize drilling operation, stuck pipe remains account for at least 25% of the nonproductive time (NPT) according to an industry report (Muqeem et al., 2012), stuck pipe might result in drill string to be cut or borehole abandoned in the worse scenario (Magana-Mora et al., 2019). The earlier a stuck pipe incident can be identified, and timely remedial action can be taken, the higher the probability of avoiding a stuck pipe or successfully freeing the pipe. (Jahanbakhshi et al., 2012). Given its severity, it is of great importance to develop a method to detect early sign of impending stuck pipe, and provide guidance for subsequent remedial measures.

Conventionally, existing methods for stuck pipe prediction can be roughly grouped into two algorithm categories: Machine Learning (ML)and Deep Learning (DL). (Jahanbakhshi et al., 2012), (Al-Baiyat and Heinze, 2012), and (Abbas et al., 2019) developed both support vector machine (SVM) and artificial neural network (ANN) approaches to predict the occurrence of stuck pipe, their studied revealed that SVM generates more promising results than ANN. Besides, decision tree (DT) is also an available algorithm in the field of stuck pipe prediction (Alshaikh et al., 2019). (Magana-Mora et al., 2019) suggested using analysis of variance (ANOVA) to analyze various drilling parameters, they sorted the resulting F-values to select features and exclude parameters with low correlation to stuck pipe incidents, and presented a robust and fast classification model based on random forest (RF) to improve the accuracy of prediction. Regarding deep learning methods, ( Tsuchihashi et al., 2021) assumed that anomalous drilling data show the same patterns, they utilized field drilling data and presented a supervised learning method called 3D convolutional neural network (CNN) modeling for anomalous drilling data to capture the early sign of stuck pipe. Gradient-weighted class activation map (grad-CAM) is adopted to provide physical explanation for black box algorithm. However, the model has not received sufficient training due to the drawback of supervised learning with limited training data. Besides, the causes of stuck pipe incidents during on-site drilling are diverse, the early signs of stuck pipe show different data patterns depending on the mechanisms of stuck pipe, these two factors together result in weak regression in the data being used. These could be the reason why exact stuck pipe sign can hardly be accurately identified and labeled. In order to overcome the limitation of supervised learning method, (Mopuri et al., 2022) employed an unsupervised deep learning tool named autoencoder (AE) on long short term memory (LSTM) modeling for normal drilling activity to detect early sign of abnormalities. The explanation of prediction is obtained by establishing the reconstruction error analysis of individual drilling parameter in the output of model. (Inoue et al., 2022) combined both supervised and unsupervised machine learning approaches. Specifically, LSTM-AE model and probability mixture model are the output of the reconstruction value and the prediction value, respectively. The stuck pipe probability is calculated by measuring the Mahalanobis distance between prediction values and measured values. (Siqueira et al. 2024) developed an expert system based on a feature engineering and fuzzy logic approach to prevent stuck pipe incidents.

For the stuck pipe prediction task, the duration of the input sequence is expected as long as possible to better describe the data patterns under various drilling conditions during the drilling process. However, such long sequence input data poses serious challenges to model's modeling and inference capabilities. RNN-based models with recurrent structure, which significantly increased computational complexity when dealing with long input drilling sequences, it also prohibits parallel computation due to their inherently sequential nature (Song et al., 2017). When dealing with stuck pipe prediction tasks, this structure also encounters issues related to error accumulation when predicting long sequences and the vanishing and exploding gradient problem. Consequently, the training time becomes excessively long, which is difficult to accept for real-time prediction. Despite the emergence of various variants such as LSTM (Hochreiter and Schmidhuber, 1997) and GRU (Cho et al., 2014), the fundamental constraint imposed by recurrent structures and the memory bottleneck in sequence computations still remain. Besides, the above-mentioned deep learning methods assume that the 'Stuck' events will occur within 0-6 minutes of the current processing segment. Though, this approach is somewhat questionable in real-time on-site stuck pipe prediction, as drilling engineers do not know when the early sign actually begin ( Tsuchihashi et al., 2021).

Recently, researchers have witnessed the tremendous success of the Transformer architecture in the field of natural language processing (NLP) and computer vision (CV) (Vaswani et al., 2017). This approach, based solely on the attention mechanism and excluding deep learning models with recurrent and convolutional structures, has demonstrated state-of-the-art performance in practical applications. This success is attributed to the Transformer's ability to model long sequences and make multi-step predictions. However, (Child et al., 2019) complained that the computation and memory complexity of the canonical Transformer self-attention mechanism grow quadratically with the length of the input sequence, which make it challenging to deploy them for long sequence problems with fine granularity. They noticed the potential sparsity of the self-attention mechanism and designed Sparse Transformer without significantly sacrificing the performance of model. (Li et al., 2020) introduced a novel convolutional self-attention mechanism, known as LogSparse self-attention, which efficiently captures local features within the input sequence. This mechanism enables each cell to attend to its previous one by an exponential step size, thus effectively reducing computational complexity. Henceforward, there have been emerged numerous variants of the Transformer model designed for long sequence time series forecasting ((Zhou et al., 2023);(Wu et al., 2022);(Zhou et al., 2022)). Transformer have been used in forecasting (Zhou et al., 2022), anomaly detection (Xu et al., 2022), and classification ((Zerveas et al., 2021), (Yang et al., 2021)).

In order to address the shortcomings of the aforementioned study, we utilize the Crossformer for capturing the long-term dependencies among parameters to detect early signs of stuck events in the drilling process, visualizing explanations of stuck risk during each time step is also employed to support the decision-making of drilling engineers. The results indicates that our method achieves superior predictive performance for impending stuck events during the drilling process. The contributions of this paper are summarized as follows:

(1) A preliminary development of the first publicly available time series dataset for the Volve field is initiated, providing data availability and serving as a benchmark dataset for future time-series-based stuck pipe prediction research.

(2) The sliding window technique is integrated into Crossformer to obtain longer output and display, addressing the limitation of canonical Crossformer.

(3) To address the issue of insufficient modeling and predictive capabilities of existing models for long sequence time series of drilling data, the improved Crossformer is introduced to capture the long-term dependencies of data during drilling process. To the best of our knowledge, this is the first application of Transformer-based models in the drilling and even petroleum engineering, offering a novel approach for stuck pipe prediction.

(4) We propose the genuinely dynamic warning threshold method in the field of stuck pipe prediction. To address the issue of accurately establishing warning threshold in previous research, dynamic warning thresholds are established based on the probability density function of reconstruction errors from normal drilling data，

This approach allows us to obtain warning threshold that adapt to variations in different drilling conditions.

(5) Leveraging the capability and characteristics of Crossformer for multi-step prediction, the relative reconstruction error between the predictions after $\tau$ steps and the ground truth at the current time is regard as the stuck risk after $\tau$ steps. This approach provides an early assessment of the risk of a stuck pipe for $\tau$ steps ahead of the current time, effectively extending the warning time.

(6) We uncover the algorithmic and structural advantages of multi-step prediction over single-step prediction methods for addressing the stuck pipe problem.

## 2. Dataset

2.1 Data description

The Volve field dataset was made public by Equinor in 2018 (Equinor, 2018), providing researchers in the oil and gas sector with valuable data resources and enabling the establishment of a public dataset benchmark. Typical raw drilling data consist of sensor data measurement while drilling (MWD), it comprises different well profiles, such as vertical, horizontal and inclined wells, some well sections also include casing and completion data. The dataset comprises both time-based data and depth-based data, prior to our research, (Tunkiel et al., 2021) utilized the depth-based data for inclination prediction. Our study employs the time-based data from the Volve drilling dataset for stuck pipe prediction, i.e., it represents data with time on rows and drilling attributes on columns. Time-based data, as compared to depth-based data, contains more attribute information (Tunkiel et al., 2020), moreover, within the same attribute, time-based data provides a more specific description of data patterns for individual attributes because of its finer data granularity.

2.2 Data pre-processing

The time-based benchmark is established first, the international standard time format of raw drilling temporal data is converted to a time format supported by Pandas library. Due to the presence of casing and completion data in some of the raw drilling data, as well as anomalous data where stuck pipe incidents were not resolved for a long time, such time segments are not suitable for stuck pipe prediction research. Additionally, the data suffers from issues such as missing important attributes and excessively large time intervals between sequences. These segments have been manually removed from the time series. The raw drilling data also contains various time intervals (i.e., 4 seconds, 15 seconds, 2 minutes), assuming the drilling parameters are continuously changing, to standardize the data, we perform linear interpolation and resampling on the data we used and segment it into uniform interval with a resolution of 4 seconds.

Regarding the selection of drilling parameters, the parameters that can be directly controlled by the driller are considered important (Salminen et al., 2017), such as mud flow in, weight on bit, rotary speed. The parameters that can reflect the downhole drilling conditions are also considered important: hookload, rate of penetration, standpipe pressure and torque (Magana-Mora et al., 2019). To enhance the visualization effect, information regarding hole depth (MD), bit depth (MD) and block position is also utilized to display drilling depth and depth increments. The aforementioned parameters are attributes shared by the Volve WITSML (wellsite information transfer standard markup language) time series dataset. Due to certain parameters being irrelevant to our research, having a significant number of missing values, or having similar attributes that can serve as substitutes, these attributes are excluded from consideration. After removing the apparent outliers from the recorded data, we retained the required dataset TSPP (time-series stuck pipe prediction). The source and details of the dataset used in this study are displayed in Table 1, where TSPP1 and its derived series did not experience stuck pipe incidents, which is considered as 'normal' drilling datasets, the remaining datasets all contain data related to stuck pipe incidents. TSPP1-15s is obtained by resampling TSPP1 at 15-second intervals, TSPP1 is a subset of TSPP1-long, and TSPP1-long exhibits stronger parameter volatility compared to TSPP1. The details of the considered parameters are shown in Table 2. Our dataset is made publicly available at: (https://github.com/promiseeee/Time-series-stuck-pipe-prediction).

Table 1 The source of data and its description.

| Dataset | Folder | Well | Start date | End date |
|---|---|---|---|---|
| TSPP1 | Norway-Statoil | F-12 | 2007-07-06 19:23:52 | 2007-07-07 18:51:48 |
| TSPP1-15s | Norway-Statoil | F-12 | 2007-07-06 19:23:45 | 2007-07-07 18:51:45 |
| TSPP1-long | Norway-Statoil | F-12 | 2007-07-06 02:34:40 | 2007-07-08 20:18:48 |
| TSPP2 | Norway-StatoilHydro | F-4 | 2007-10-23 08:07:08 | 2007-10-23 17:33:40 |
| TSPP3 | Norway-NA | F-9 A | 2009-07-03 18:04:32 | 2009-07-04 08:00:00 |
| TSPP4 | Norway-NA | F-1 | 2007-12-01 10:03:12 | 2007-12-01 22:00:00 |

Table 2 Description of the parameters considered by the proposed method.

| Type | Parameter | Unit |
|---|---|---|
| Depth parameters | Hole depth (MD) | m |
| | Bit depth (MD) | m |
| | Block position | m |
| Drilling parameters | Torque | kN.m |
| | Hookload | kkgf |
| | Rotary speed | RPM |
| | Standpipe pressure (SPP) | kPa |
| | Mud flow in | L/min |
| | Weight on bit (WOB) | kkgf |
| | Rate of penetration (ROP) | m/h |

After performing normalization on each dimension of the parameters, the data is transformed to have a mean of 0 and a standard deviation of 1. The processed data is split into three parts: train, validation, and test, with a ratio of 0.7:0.1:0.2, respectively. The labeling method is similar to (Mopuri et al., 2022), as for a dataset containing stuck pipe case, the model is trained and validated using 'normal' drilling data, the test set should include all the early signs and abnormal data related to stuck pipe incident, as shown in Fig. 1. Where $\Lambda$ is the length of test set or the number of sliding window steps, $\tau$ is the prediction step size. In our study, the 'stuck' portion of the test set is shaded in light red.

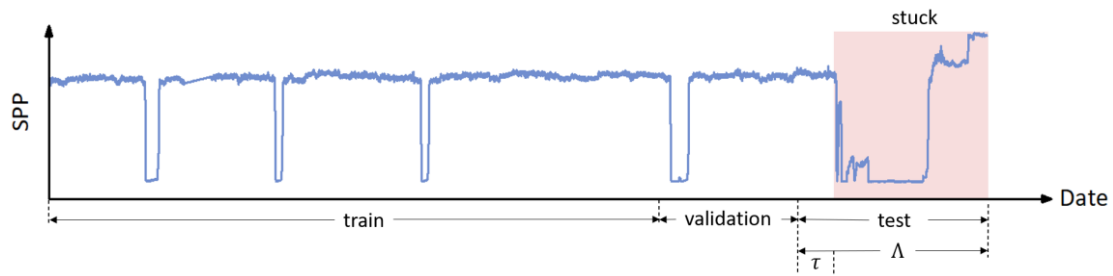

Fig. 1. Example of data division using SPP, the early signs of stuck pipe are located just before the 'stuck' portion.

# 3. Model choose/design

The sequential nature of RNNs severely limits their parallelization, making it challenging to perform large-scale learning, the recurrent structure of RNNs can also lead to a decrease in inference speed. Furthermore, RNN-based models have consistently struggled with addressing long-term dependency issues and often require the assistance of attention mechanisms. Additionally, there is an extreme imbalance between the labels 'normal' and 'stuck' in drilling data, the issue of data imbalance poses the awkward dilemma of oversampling or undersampling in supervised learning, resulting in the loss of data continuity properties. Due to the powerful long sequence modeling capability of Transformer-based model, the data used is non-overlapping and does not require partitioning, allowing modeling of dependencies without regard to their distance in the input or output sequences (Vaswani et al., 2017), the Transformer-based model also has superior parallel computing capability, thanks to the introduction of the multi-head attention mechanism.

Existing Transformer models embed data points into a vector at the same time step, and the model primarily focuses on capturing temporal dependency within the sequence. Previous researchers have made significant advancements based on the sparsity of the attention layer ((Li et al., 2020); (Zhou et al., 2023); (Wu et al., 2022)). However, the fragmentary nature of the attention layer has not been fully utilized, as illustrated in Fig. 2. Bourgoyne and Young's ROP model reveals a coupling relationship among ROP, WOB, Rotary Speed, and hydraulic factors (Bourgoyne and Young, 1974). Due to the value of parameter at a single time step can provide limited information (Zhang and Yan, 2023), it is essential to unearth the cross-dimension dependencies within the drilling data.

In this section, we introduce the architecture of Crossformer model (Zhang and Yan, 2023), which is one of the variants for time series forecasting based on canonical Transformer model. In essence, Crossformer enhances the modeling capability of the Transformer architecture by allowing it to effectively model both temporal and cross-dimensional dependencies, making it well-suited for a broader range of applications that require understanding complex relationships within drilling data. By incorporating cross-dimensional relationships, Crossformer can potentially improve the performance of model on tasks that involve complex patterns or interactions among various parameters.

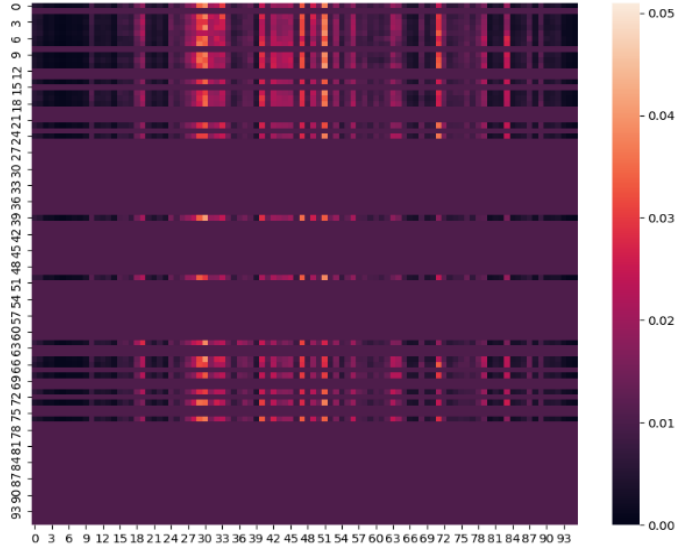

Fig. 2. Self-attention scores based on Transformer is created on TSPP1, where brighter colors indicate higher Self-attention scores. This illustrates that the drilling data tends to be segmented.

3.1 Dimension-segment-wise embedding and positional encoding

In contrast to conventional Transformer-based models, which embed the data points at each time step into a vector, the dimension-segment-wise data partitioning method embedding a series segment from a single dimension into a vector:

$$\begin{aligned} X_{1:T} &= \left\{ X_{i,d}^{(s)} \middle| 1 \le i \le \frac{T}{L_{seg}}, 1 \le d \le D \right\} \\ X_{i,d}^{(s)} &= \left\{ X_{t,d} \middle| (i-1) \times L_{seg} < t \le i \times L_{seg} \right\} \end{aligned} \quad (1)$$

Where $X_{1:T} \in \mathbb{R}^{T \times D}$ is the input of Crossformer, $D$ is the dimension of input data, $X_{i,d}^{(s)} \in \mathbb{R}^{L_{seg}}$ is the $i$-th segment in dimension $d$ with length $L_{seg}$. Subsequently, each segment is embedded into a vector using linear projection along with the addition of a position embedding, sliding windows technology is integrated to address the output and display issue:

$$h_{i,d} = Slide(\boldsymbol{E} X_{i,d}^{(s)} + \boldsymbol{E}_{i,d}^{pos}) \quad (2)$$

Where $\boldsymbol{E} \in \mathbb{R}^{d_{model} \times L_{seg}}$ denotes the learnable projection, $\mathbb{R}^{d_{model}}$ denotes the dimension of model, $\boldsymbol{E}_{i,d}^{pos} \in \mathbb{R}^{d_{model}}$ denotes the learnable position embedding for position $(i, d)$, $Slide$ denotes the sliding window operation. After incorporating the embedded data with a sliding window, a 2D vector array $\boldsymbol{H} = \left\{ h_{i,d} \middle| 1 \le i \le \frac{T}{L_{seg}}, 1 \le d \le D \right\}$ is obtained, where each $h_{i,d}$ represents a univariate time series segment. Then the embedded input and positional encoding are concatenated together and fed into the encoder layers by a sliding window. The canonical Crossformer only outputs predictions for $\tau$ time steps ahead, however, the transition from normal drilling conditions to becoming stuck typically takes several minutes or even tens of minutes. Therefore, displaying outputs for only $\tau$ steps are not suitable for the task of stuck pipe prediction, the introduction of the sliding window technique can allow the display of the entire test dataset.

3.2 Two-stage attention

Attention mechanism has achieved remarkable success in sequence modeling, which enables modeling the dependency of sequences without considering their distance in the input or output (Bahdanau et al., 2014), this approach get rid of recurrent and convolutional structure that rely on inherent sequential properties. Multi-head attention enables the model to simultaneously focus on various aspects of information from distinct representation subspaces and positions:

$$head_i = Attention(QW_i^Q, KW_i^K, VW_i^V)$$
$$MSA(Q, K, V) = Concat(head_1, head_2, \cdots, head_h)W^O \quad (3)$$

Where $W_i^Q, W_i^K, W_i^V$ are parameter matrices for linear projection. $Q \in \mathbb{R}^{L_{seg} \times D}$, $K \in \mathbb{R}^{L_{seg} \times D}, V \in \mathbb{R}^{L_{seg} \times D}, W^O \in \mathbb{R}^{hL_{seg} \times D}$ serve as queries, keys and values, $Concat$ denotes the concatenation operation.

For drilling engineering that exhibits data with long periodicity, the capability to extract long-range dependency is crucial in such cases. However, applying self-attention to a 2D array results in quadratic computational complexity, which is unbearable for long time series, the proposed Two-Stage Attention ($TSA$) with router mechanism is used to address this efficiency issue and capture cross-time and cross-dimension dependencies. In the cross-time stage, since the input vectors are obtained through dimension-segment-wise embedding, canonical multi-head self-attention ($MSA$) is directly applied to each dimension:

$$\widehat{\mathbf{Z}}_{:,d}^{time} = LayerNorm\left(\mathbf{Z}_{:,d} + MSA^{time}(\mathbf{Z}_{:,d}, \mathbf{Z}_{:,d}, \mathbf{Z}_{:,d})\right)$$
$$\mathbf{Z}^{time} = LayerNorm\left(\widehat{\mathbf{Z}}^{time} + MLP(\widehat{\mathbf{Z}}^{time})\right) \quad (4)$$

$\mathbf{Z}_{i,:}$ denotes the vectors of all dimensions at time step $i$, $\mathbf{Z}_{:,d}$ denotes all time steps in dimension $d$. $\mathbf{Z}$ here can be the output of DSW embedding or lower TSA layers, $LayerNorm$ denotes layer normalization, $MSA(Q, K, V)$ denotes the multi-head self-attention, $MLP$ denotes a multi-layer feedforward network, $\widehat{\mathbf{Z}}^{time}$ and $\mathbf{Z}^{time}$ denote the output of the $MSA$ and $MLP$, respectively. The cross-time dependency of segments is captured in $Z^{time}$.

In the cross-dimension stage, a router mechanism is proposed to address the limitation resulting from quadratic computational complexity. These routers initially aggregate messages from all dimensions by employing routers as query within the MSA framework, with vectors from all dimensions serving as key and value. Subsequently, these routers distribute the gathered messages among dimensions, by using vectors of dimensions as queries and the aggregated messages as keys and values, the aforementioned process can be described as follows:

$$\begin{aligned} \boldsymbol{B}_{i,:} &= MSA_1^{dim}(\boldsymbol{R}_{i,:}, \boldsymbol{Z}_{i,:}^{time}, \boldsymbol{Z}_{i,:}^{time}), 1 \leq i \leq T \\ \overline{\boldsymbol{Z}}_{i,:}^{dim} &= MSA_2^{dim}(\boldsymbol{Z}_{i,:}^{time}, \boldsymbol{B}_{i,:}, \boldsymbol{B}_{i,:}), 1 \leq i \leq T \\ \widehat{\boldsymbol{Z}}^{dim} &= LayerNorm(\boldsymbol{Z}^{time} + \overline{\boldsymbol{Z}}^{dim}) \\ \boldsymbol{Z}^{dim} &= LayerNorm(\widehat{\boldsymbol{Z}}^{dim} + MLP(\widehat{\boldsymbol{Z}}^{dim})) \end{aligned} \quad (5)$$

Where $\boldsymbol{R} \in \mathbb{R}^{d_{model}}$ (c is a constant, $c \ll D$) denotes the learnable vector array for each time step $i$ as routers, $\boldsymbol{B} \in \mathbb{R}^{T \times c \times d_{model}}$ denotes the aggregated messages from all dimensions. $\overline{\boldsymbol{Z}}^{dim}$ denotes output of the router mechanism, $\widehat{\boldsymbol{Z}}^{dim}, \boldsymbol{Z}^{dim}$ denote the output of skip connection and $MLP$, respectively. Simultaneous Eq.4 and Eq.5, two-stages attention model is derived as follows:

$$\boldsymbol{Y} = \boldsymbol{Z}^{dim} = TSA(\boldsymbol{Z}) \quad (6)$$

Where $\boldsymbol{Y}, \boldsymbol{Z} \in \mathbb{R}^{T \times D \times d_{model}}$ denote the input and output vector array of $TSA$ layer, respectively. Both cross-time and cross-dimension dependencies are captured in $\boldsymbol{Y}$.

3.3 Hierarchical encoder-decoder

Encoder-decoder structures have shown their prowess in sequence modeling (Sutskever et al., 2014), especially in Transformer models. The hierarchical encoder-decoder (HED) is used to capture the key features and relevant information of the input sequence, compressing it into a fixed-length vector, and outputting at different scales. The encoding process can be described as $\boldsymbol{Z}^{enc,l} = Encoder(\boldsymbol{Z}^{enc,l-1})$:

$$\begin{cases} l = 1: \widehat{\boldsymbol{Z}}^{enc,l} = \boldsymbol{H} \\ l > 1: \widehat{\boldsymbol{Z}}_{i,d}^{enc,l} = \boldsymbol{M}[\boldsymbol{Z}_{2i-1,d}^{enc,l-1} \cdot \boldsymbol{Z}_{2i,d}^{enc,l-1}], 1 < i < \frac{L_l - 1}{2}, 1 \leq d \leq D \\ \boldsymbol{Z}^{enc,l} = TSA(\widehat{\boldsymbol{Z}}^{enc,l}) \end{cases} \quad (7)$$

Where $\boldsymbol{H}$ denotes the 2D array output of DSW embedding, $\boldsymbol{Z}^{enc,l}$ denotes the output of the $l$-th encoder layer, $\boldsymbol{M} \in \mathbb{R}^{d_{model} \times 2d_{model}}$ denotes a learnable matrix for segment merging, $[\cdot]$ denotes the concatenation operation, $L_l - 1$ denotes the number of segments in each dimension in layer $l-1$, $\hat{\mathbf{Z}}^{enc,l}$ denotes the array after segment merging in the $l$-th layer, $\mathbf{Z}^{enc,0}, \mathbf{Z}^{enc,1}, \cdots, \mathbf{Z}^{enc,N}$ represent the N+1 outputs of the encoder. The obtained feature array $\mathbf{Z}^{enc,l}$ is fed into the decoder to obtain a $l$-layer 2D decoded array, which can be depicted as $\mathbf{Z}^{dec,l} = Decoder(\mathbf{Z}^{dec,l-1}, \mathbf{Z}^{enc,l})$:

$$\begin{cases} l = 0: \widetilde{\mathbf{Z}}^{enc,l} = TSA(\mathbf{E}^{(dec)}) \\ l > 0: \widetilde{\mathbf{Z}}^{enc,l} = TSA(\mathbf{E}^{(dec,l-1)}) \end{cases}$$
$$\overline{\mathbf{Z}}_{:,d}^{dec.l} = MSA(\widetilde{\mathbf{Z}}_{:,d}^{enc,l}, \mathbf{Z}_{:,d}^{enc,l}, \mathbf{Z}_{:,d}^{enc,l}), 1 \leq d \leq D$$
$$\widehat{\mathbf{Z}}^{dec,l} = LayerNorm\left(\widetilde{\mathbf{Z}}^{enc,l} + \overline{\mathbf{Z}}^{dec,l}\right)$$
$$\mathbf{Z}^{dec,l} = LayerNorm\left(\widehat{\mathbf{Z}}^{dec,l} + MLP(\widehat{\mathbf{Z}}^{dec,l})\right)$$
(8)

where $\mathbf{E}^{(dec)} \in \mathbb{R}^{\frac{\tau}{L_{seg}} \times D \times d_{model}}$ denotes the learnable position embedding for decoder, $\widetilde{\mathbf{Z}}^{enc,l}$ is the output of $TSA$, $\overline{\mathbf{Z}}_{:,d}^{dec.l}$ denotes the output of $MSA$, the $MSA$ layer establishes a link between the encoder and decoder by utilizing $\widetilde{\mathbf{Z}}_{:,d}^{enc,l}$ as the query and $\mathbf{Z}_{:,d}^{enc,l}$ as both the key and value, $\widehat{\mathbf{Z}}^{dec,l}$ and $\mathbf{Z}^{dec,l}$ denote the output of skip connection and $MLP$, respectively. $\mathbf{Z}^{dec,0}, \mathbf{Z}^{dec,1}, \cdots, \mathbf{Z}^{dec,N}$ represent the N+1 outputs of the decoder, N is the number of encoder layers.

Linear projection is employed on the output of each layer to generate predictions, for the $\lambda$-th step prediction, the predictions of these layer within the $\lambda$-th sliding window are then aggregated through summation to produce the prediction:

$$for\ l = 0, \cdots N: x_{i,d}^{(s),l} = \mathbf{W}^l \mathbf{Z}_{i,d}^{dec,l}$$
$$x_{T+\lambda+1:T+\lambda+\tau}^{pred,l} = \left\{ x_{i,d}^{(s),l} \middle| 1 \leq i \leq \frac{\tau}{L_{seg}}, 1 \leq d \leq D, 0 \leq \lambda \leq \Lambda \right\}$$
$$\mathbf{P}_\lambda = x_{T+\lambda+1:T+\lambda+\tau}^{pred} = \sum_{l=0}^{N} x_{T+\lambda+1:T+\lambda+\tau}^{pred,l}, 0 \leq \lambda \leq \Lambda$$
(9)

where $\mathbf{W}^l \in \mathbb{R}^{L_{seg} \times d_{model}}$ is a learnable matrix to project a vector to a time series segment. $x_{i,d}^{(s),l} \in \mathbb{R}^{L_{seg}}$ denotes the prediction of $i$-th segment in dimension $d$, $x_{T+\lambda+1:T+\lambda+\tau}^{pred,l}$ denotes the prediction of all segments at layer $l$ within the $\lambda$-th step sliding window, the prediction $\mathbf{P}_\lambda$ in the $\lambda$-th step sliding window is obtained by summing up the predictions from all the layers.

For each $\mathbf{P}_\lambda$ obtained at time step $\lambda$, the sliding window is adopted to display the prediction $\mathbf{P}_{\lambda+1}$ for the next time step $(\lambda + 1)$, the prediction results $\mathbf{P}_{T+1:\Lambda+\tau}$ for all time steps is obtained by concatenating the results of $\Lambda + \tau$ steps:

$$\begin{aligned}
\boldsymbol{P}_{\lambda+1} &= Slide(\boldsymbol{P}_\lambda) = x^{pred}_{T+\lambda+2:T+\lambda+\tau+1}, 0 \leq \lambda \leq \Lambda \\
\boldsymbol{P}_{T+1:T+\Lambda+\tau} &= Concat\left(x^{pred}_{T+1:T+\tau}, x^{pred}_{T+\tau+1}, x^{pred}_{T+\tau+2}, \cdots, x^{pred}_{T+\tau+\Lambda}\right)
\end{aligned} \quad (10)$$

Assuming the output generated by the previous sliding window is $x^{pred}_{T+\lambda+1:T+\lambda+\tau}$, when the step size of sliding window is set to 1, the output of the next sliding window is $x^{pred}_{T+\lambda+2:T+\lambda+\tau+1}$, where only $x^{pred}_{T+\lambda+\tau+1}$ is the latest output value, $x^{pred}_{T+\lambda+2:T+\lambda+\tau}$ has already been predicted separately $(\lambda-1:1)$ times in previous windows, only the latest output, i.e., $x^{pred}_{T+\lambda+\tau+1}$, is selected and concatenated to form $\boldsymbol{P}_{T+1:T+\Lambda+\tau}$. Since $x^{pred}_{T+1:T+\tau}$ is the output obtained in the first step, they are already the latest values at the initial moment, consequently, the first $\tau$ outputs are directly selected to be used as the subset of $\boldsymbol{P}_{T+1:T+\Lambda+\tau}$. The sketch of our model is shown in Fig. 3.

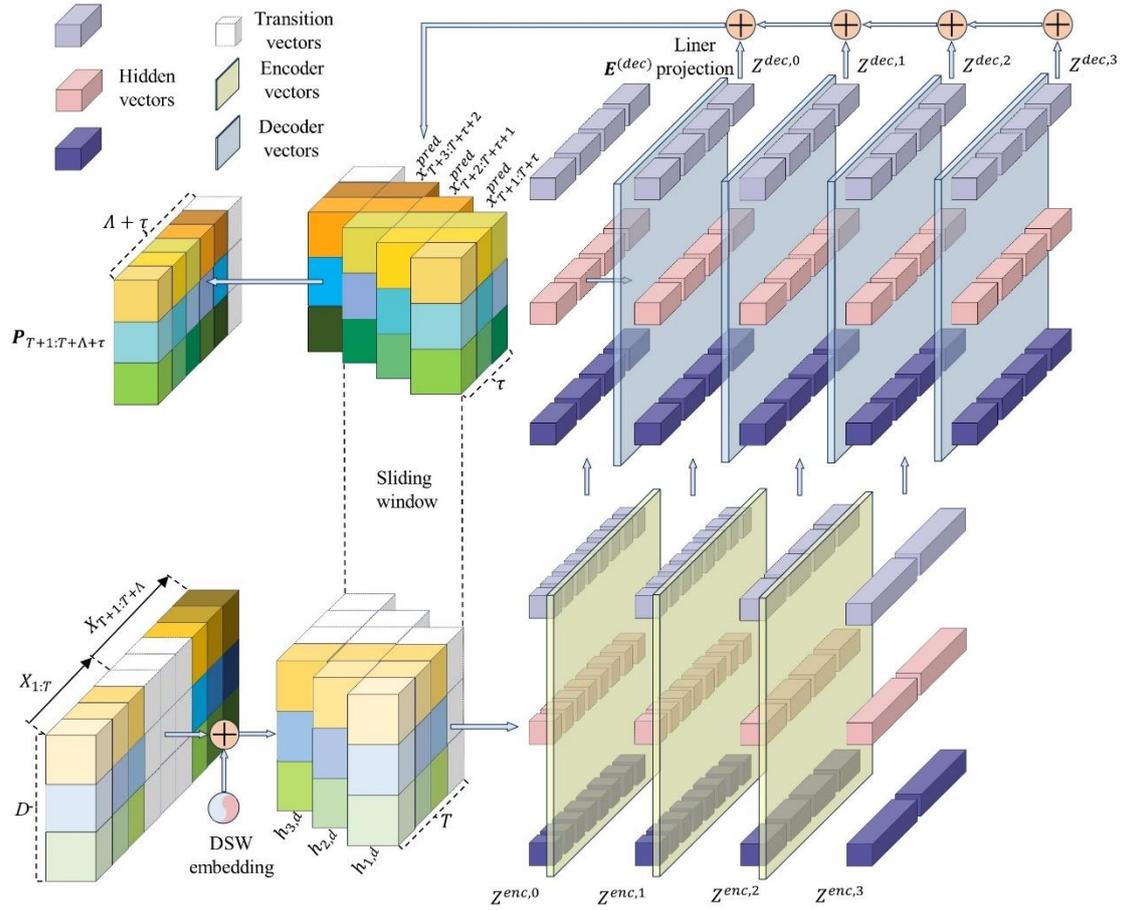

Fig. 3. Model diagram description for a 3-layer decoder. The drilling data is embedded using dimension-segment-wise embedding after normalization. The positionally encoded data is fed into the encoder module using a sliding window with step size of 1, the cross-time and cross-dimension dependencies are captured by the two-stage attention module, the decoder outputs predictions of length $\tau$ for each step of the sliding window. These predictions are concatenated to form the final model output $\boldsymbol{P}_{T+1:T+\Lambda+\tau}$. Where transition vectors represent segments of vectors omitted due to diagram constraints. $\boldsymbol{Z}^{enc,l}$ and $\boldsymbol{Z}^{dec,l}$ represent the outputs of the $l$-th layer of the encoder and decoder, respectively. $X_{1:T+\Lambda}$ denotes the drilling data of length $(T+\Lambda)$ and dimension $D$, $h_{i,d}$ denotes the input data for the encoder with a length of $T$, $x^{pred}_{T+\lambda+1:T+\lambda+\tau}$ denotes the output data of the decoder with a length of $\tau$. $X_{T+1:T+\Lambda}$ and $P_{T+1:T+\Lambda+\tau}$ are used as ground truth and predictions to the subsequent establishment of risk, respectively.

## 4. Methodology

4.1. Evaluation method

To evaluate the predictive performance of model, mean absolute error ($MAE$) and mean squared error ($MSE$) are introduced as evaluation metrics, given by the following:

$$MAE = \frac{1}{n}\sum_{i=0}^{n}|y_i - \hat{y}_i|$$
$$MSE = \frac{1}{n}\sum_{i=0}^{n}(y_i - \hat{y}_i)^2 \quad (11)$$

Where $n$ is the number of predicting samples, $y_i$ and $\hat{y}_i$ are the $i$-th ground truth and prediction of model, respectively. All experiments are to be repeated at least five times, excluding the maximum and minimum values, the ultimate $MAE\ and\ MSE$ is obtained by calculating the average of the remaining three results.

In order to balance the numerical values across different parameters, relative reconstruction error is utilized to assess the risk of stuck pipe at each time step within each parameter, the sum of the relative reconstruction error for each drilling parameter yields the relative reconstruction error at each time step, denoted as $Risk_{T:T+\Lambda+\tau}$:

$$\boldsymbol{R}_{T:T+\Lambda+\tau;d} = \begin{cases} \frac{\left|x^{truth}_{T-\tau:T;d} - x^{pred}_{T+1:T+\tau;d}\right|}{x^{truth}_{T-\tau:T;d}} \times 100\%, \lambda = 0 \\ 0, \quad \left|x^{truth}_{T+\lambda+1;d}\right|\ or\ \left|x^{truth}_{T-\tau:T;d}\right| \approx 0 \\ \frac{\left|x^{truth}_{T+\lambda+1;d} - x^{pred}_{T+\lambda+\tau;d}\right|}{x^{truth}_{T+\lambda+1;d}} \times 100\%, 0 < \lambda \leq \Lambda, \left|x^{truth}_{T+1;d}\right| > 0 \end{cases}, 3 \leq d \leq D \quad (12)$$

$$Risk_{T:T+\Lambda+\tau} = \sum_{d=3}^{D} Norm(\boldsymbol{R}_{T:T+\Lambda+\tau;d})$$

Where $x^{truth}_{;d}$ denotes the ground truth for all time steps in dimension $d$, $x^{pred}_{;d}$ denotes the predictions for all time steps in dimension $d$. $\boldsymbol{R}_{T:T+\Lambda+\tau;d}$ denotes the relative reconstruction error for $(T:T+\Lambda+\tau)$ time steps in dimension $d$. Due to the introduction of depth parameters causing the distortion of $Risk$, as seen in Fig. 7 and 8, they are excluded when calculating $Risk$. Furthermore, due to exceptionally low ground truth values for some of the data (mud flow in), which cause the $\boldsymbol{R}_{T:T+\Lambda+\tau;d}$ values to be abnormally high, these values are set to zero to minimize the impact of normalization on normal data.

We use $x^{truth}_{T+\lambda+1;d}$ instead of $x^{truth}_{T+\lambda+\tau;d}$ to calculate the relative reconstruction error, because when use $x^{truth}_{T+\lambda+\tau;d}$ for relative reconstruction error calculation, one can only make the calculation when $x^{truth}_{T+\lambda+\tau;d}$ is received at time step $(T+\lambda+\tau)$, which reduces the time available for on-site drilling engineers to implement remedial measures once the risk has been identified. Considering that the model can obtain predictions for $\tau$ steps ahead, using $x^{truth}_{T+\lambda+1;d}$ instead of $x^{truth}_{T+\lambda+\tau;d}$ allows the model to obtain $Risk$ for $\tau$ steps ahead of the current time $(T+\lambda)$. In the physical significance, this represents the relative variation rate of the upcoming value $x^{pred}_{T+\lambda+\tau;d}$ relative to the current value $x^{truth}_{T+\lambda+1;d}$, which is also valuable.

4.2 Establish warning threshold

Similar to the concept of unsupervised learning (Mopuri et al., 2022), we assume that the training data encompasses all data patterns associated with normal drilling scenarios. If the input data comes from normal drilling operations, the model is expected to make accurate predictions for the parameters, and $Risk$ is expected to be relatively small, when the model is not able to predict the upcoming data accurately, the $Risk$ output a larger value, which can be considered as early sign of a stuck pipe, in other words, the relative reconstruction error is regard as the risk of stuck pipe. Furthermore, we assume that the modeling errors of the model on normal data and the errors caused by abnormal fluctuations have a similar ratio. Based on above hypothesis, the reconstruction error $Risk$ is divided into modeling error for normal drilling data and prediction error caused by abnormal data:

$$Risk_{T:T+\Lambda+\tau} = \boldsymbol{E}^\tau + \boldsymbol{E}^p_{T:T+\Lambda+\tau} \tag{13}$$

Where $\boldsymbol{E}^\tau$ is the reconstruction error of model in predicting $\tau$ steps ahead for normal drilling data, it shows the modeling and predictive ability of model. $\boldsymbol{E}^p_{T:T+\Lambda+\tau}$ denotes the reconstruction errors caused by abnormal fluctuations in drilling parameters, it represents the variations in drilling parameters that the model can't predict. In order to make $Risk_{T:T+\Lambda+\tau}$ as close to $\boldsymbol{E}^p_{T:T+\Lambda+\tau}$ as possible, the model is expected to make accurate predictions for drilling parameters in order to achieve a small $\boldsymbol{E}^\tau$. This line of thinking is contingent upon the ability of model to predict drilling parameters accurately. Therefore, it is of great importance to assess model's predictive capability for normal drilling scenarios.

In general, when $Risk$ reaches a certain value, the model is considered unable to predict the current data accurately, therefore, this may indicate an impending stuck. The segment where the reconstruction error of model exceeds the value before the occurrence of a stuck pipe incident is considered an early sign of stuck pipe, this value is referred to warning threshold, the combined duration of prediction step size and early sign is referred to warning time, during which timely remedial actions need to be taken to prevent stuck pipe incident. Traditional alarms are designed to trigger when the measurements exceed a defined threshold. Unfortunately, the warning thresholds are expected to fluctuate under various drilling conditions, which can wreak havoc with traditional alarms. Crews are faced with a dilemma: they must decide between setting alarms with reasonable thresholds and thus dealing with many false alarms, or widening the thresholds, thereby sacrificing early detection or any detection at all (Unrau et al., 2017).

For the obtained curve $Risk$, in order to quantify the warning threshold and warning time, we assume that the normal drilling operation portion of $Risk$ in terms of frequency is a normally distribution, the warning threshold and warning time are determined by the following:

$$W_v = (1 + MSE^\tau)(\mu_{T+k_1\tau:T+k_2\tau} + 2\sigma_{T+k_1\tau:T+k_2\tau}), k_2 \gg k_1 \geq 0$$
$$W_t = t^\tau + t^p \tag{14}$$

Where $W_v$ is warning threshold, $MSE^\tau$ is the $MSE$ of the reconstruction error of model for normal drilling data with a prediction step size of $\tau$, $(1 + MSE^\tau)$ can be considered as a penalty term. For anomaly detection tasks, it is desired to assign greater weight to the detection of anomalous points, therefore, $MSE$ is chosen to construct the penalty term. $\mu_{T+k_1\tau:T+k_2\tau}$ and $\sigma_{T+k_1\tau:T+k_2\tau}$ are the mean and variance of the probability density function of $Risk$ from time $(T + k_1\tau)$ to $(T + k_2\tau)$, respectively. Due to the presence of some disturbances in normal drilling operations, we aim to differentiate between normal drilling data and abnormal data by using a 95% confidence interval. Parameters with smaller variations will primarily affect the mean, whereas parameters with larger variations will lead to increased variance, in other words, $\mu$ and $\sigma$ indicate the invariance and volatility of drilling parameters, respectively. $W_t$ is warning time, $t^\tau$ is the duration of prediction step size, $t^p$ is the duration of early sign. Since new data is continually added during drilling process, when employing this model for real-time stuck pipe prediction, the warning threshold is inherently dynamic, thus constituting a dynamic threshold, which enhances the robustness of model. When expanding Eq. 14, $MSE^\tau(\mu_{T+k_1\tau:T+k_2\tau} + 2\sigma_{T+k_1\tau:T+k_2\tau})$ and $(\mu_{T+k_1\tau:T+k_2\tau} + 2\sigma_{T+k_1\tau:T+k_2\tau})$ can be considered to correspond to $E^\tau$ and $E^p_{T:T+\Lambda+\tau}$ in Eq. 13, quantifying the modeling error and prediction error, respectively.

Different from (Tsuchihashi et al., 2021)and (Mopuri et al., 2022),which set the warning time for a fixed period, thanks to the dynamic threshold nature, when this method is applied to real-time stuck pipe prediction, the obtained threshold will change over time, for impending stuck pipe incident, the warning time can be obtained intuitively. For different wellbore sections or geological formations, we can reset $\mu$ and $\sigma$, then recalculate the threshold using data from nearby section at the current time to adapt to the changes in the drilling conditions.

4.3 Hyperparameter tuning and selection

We evaluated the impact of two hyperparameters, the number of routers (c) and segment length ($L_{seg}$) on the normal drilling dataset TSPP1. All transverse comparisons of a single parameter are conducted under the condition where all the other parameters are at their optimum. As Fig. 4(a) shows, the parameter c, which controls information bandwidth, does not have a significant impact on model performance, given our assumption that $c \ll D$, c is chosen to be 3. Segment length ($L_{seg}$) is related to the computational complexity of model ($O(\frac{D}{L_{seg}^2}T^2)$), and there is a slight increase in $MSE$ at a value of 16, as shown in Fig. 4(b), therefore, $L_{seg}$ is set to be 12.

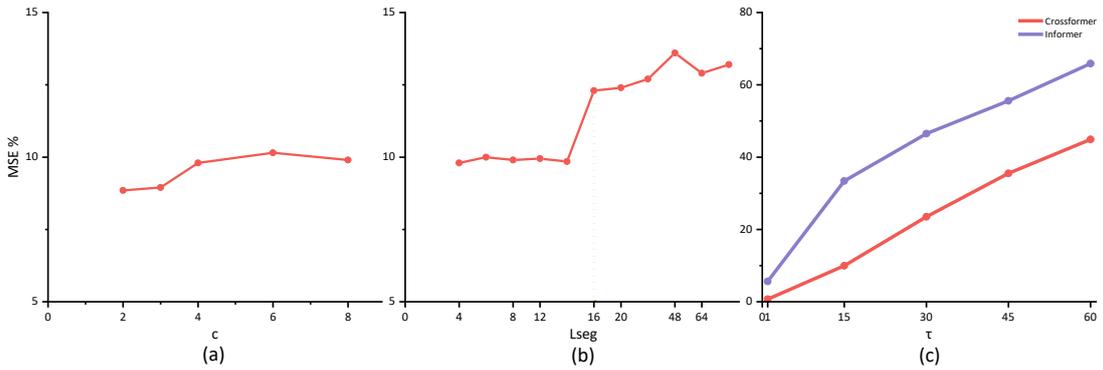

Fig. 4. (a) $MSE$ against hyperparameter number of routers c in TSA layer on TSPP1. (b) $MSE$ against hyperparameter segment length $L_{seg}$ in DSW embedding on TSPP1. (c) Comparison of the prediction performance between Informer and Crossformer at different prediction step sizes.

# 5. Results

5.1 Prediction performance comparison

The normal drilling dataset TSPP1 is used to compare the modeling capability and predictive performance between the Transformer-based model and existing DL models, such as CNN and LSTM. Since most DL models currently make prediction only one step ahead, the same prediction step size is adopted for horizontal comparison. Furthermore, the Crossformer model is utilized to investigate the impact of data granularity and input sequence length on predictive performance, the results are presented in Table 3. It is obviously that the Crossformer model exhibits significant performance advantages compared to other DL models. When utilize the Crossformer to make predictions on different datasets, finer data granularity and longer input data lengths resulted in smaller $MSE$, even when the dataset exhibited much larger fluctuations.

Table 3. The evaluation of data and model performance. When evaluating performance with Crossformer on TSPP1-15s, the prediction step size is set to 4 to align the duration of prediction.

| dataset | model | $MSE\%$ | $MAE\%$ |
|---|---|---|---|
| TSPP1 | CNN | 11.06 | 26.10 |
| TSPP1 | LSTM | 1.98 | 9.37 |
| TSPP1 | Informer | 5.63 | 19.3 |
| TSPP1 | Crossformer | 0.52 | 4.28 |
| TSPP1-15s | Crossformer | 13.78 | 22.26 |
| TSPP1-long | Crossformer | 0.13 | 1.97 |

Crossformer significantly outperforms Informer at the same prediction step size. The prediction step size $\tau$, which has a significant impact on the prediction accuracy $E^\tau$, for Informer and Crossformer models with multi-step prediction capability, we prolong the prediction step size from 1 to 60 on TSPP1 to compare modeling and prediction performance, as shown in Fig. 4(c). For $\tau > 60$, it is considered meaningless due to excessively large reconstruction error, a longer step size not only leads to a higher false alarm rate but also results in performance degradation, as shown in Table 4 for reference.

Furthermore, visual explanation for Crossformer's predictions of each parameter on dataset TSPP1 are provided, the ground truth and predictions of each parameter are shown in Fig. 5. From the graph, it is evident that the model can make accurate predictions even when there are obvious fluctuations in the parameters, with the prediction *MSE* about 0.52%. Most of the reconstruction error are observed in the prediction of depth parameters, while the model excels in predicting drilling parameters.

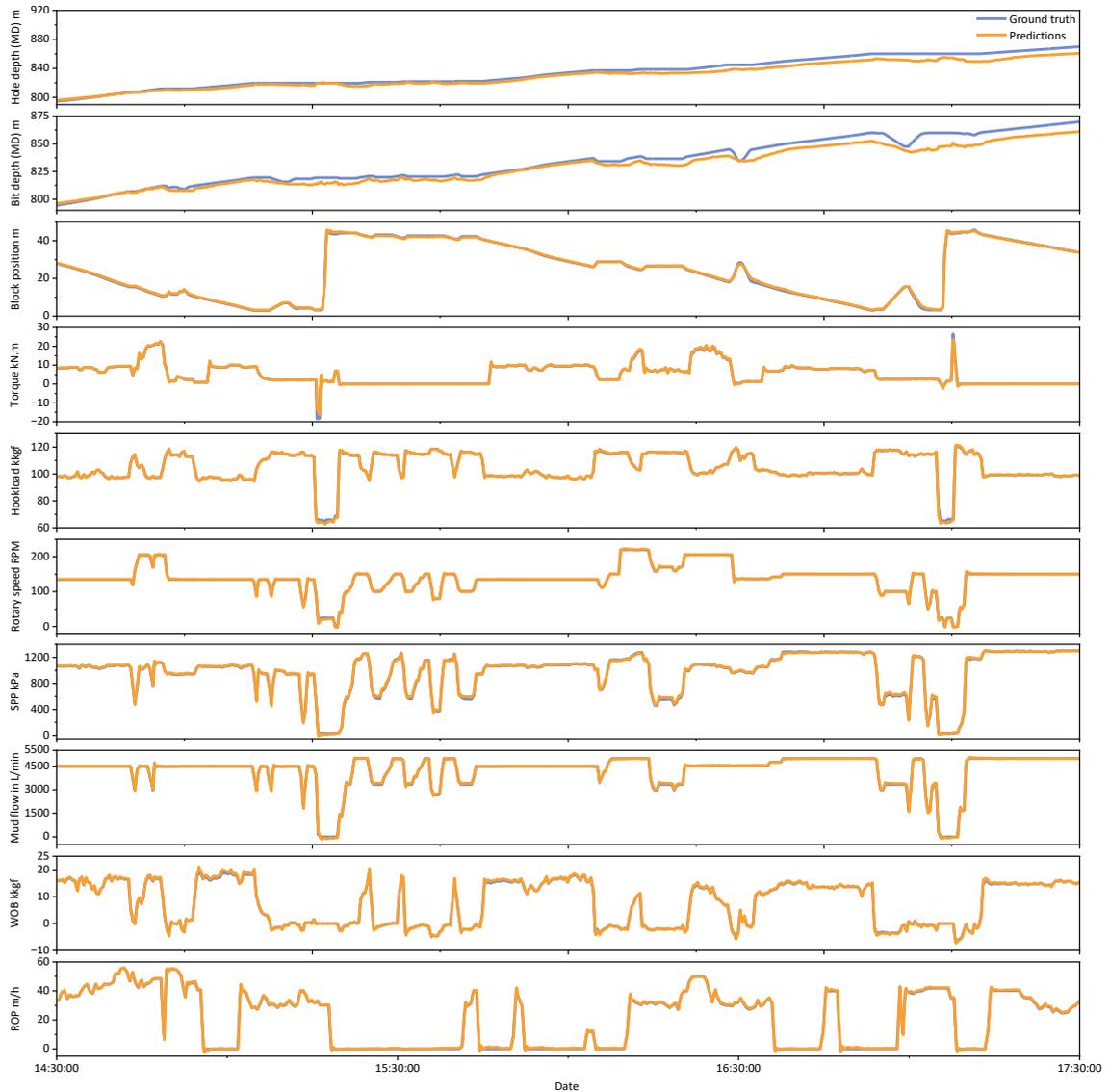

Fig. 5. When $\tau = 1$, the ground truth and predictions for various parameters are indicated in blue and yellow, respectively. Even when parameters exhibit obvious fluctuations, the model is capable of making efficient predictions.

5.2 Stuck pipe prediction

For TSPP2 containing stuck pipe incident, predictions are made at different prediction step sizes from 1 to 60 for the purpose of assessing the impact of predictive performance, as shown in Fig. 5. Although there are several fluctuations in *Risk* from 16:20 to 16:50 before stuck pipe incident, they do not directly lead to the eventual stuck pipe. Starting at 16:52, the value of *Risk* exhibits an upward trend after experiencing a fluctuation, a stuck pipe incident occurs at 16:59. The $Risk$ from 16:20 to 16:50 are selected to calculate the mean and variance of the probability density function of normal distribution for different prediction step sizes. It is evidently that all prediction steps of Crossformer successfully detect the early signs of a stuck pipe except when $\tau = 1$, the details of predictions are described in Table 4. As the prediction step size increases, the mean and variance of the probability density function gradually decrease. Since $\tau = 1$ does not detect the early sign of a stuck pipe, the warning time can't be provided.

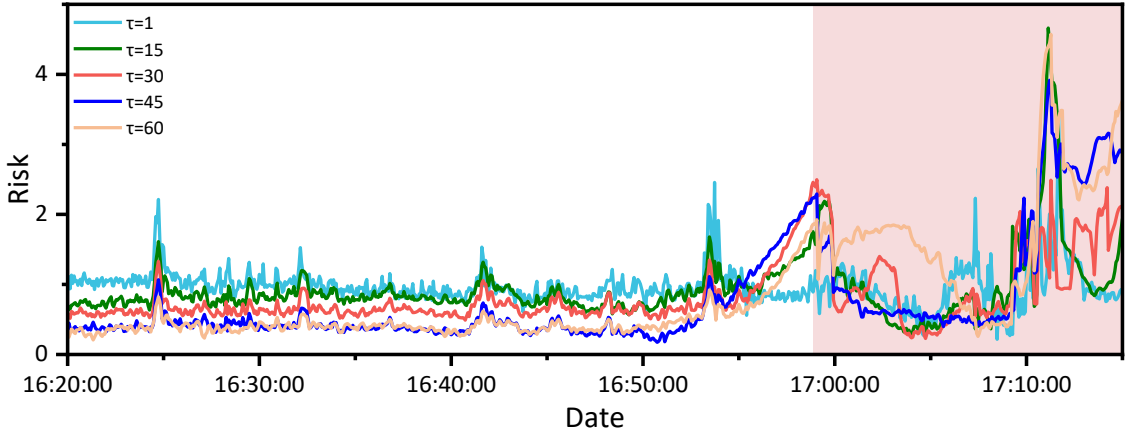

Fig. 6. The curves of $Risk$ over time under different prediction step sizes demonstrate that our model successfully detects the early signs of a stuck pipe.

Table 4. The means and variances of the probability density function for normal drilling data under different prediction step sizes, as well as the warning thresholds and warning times.

| $\tau$ | $\mu$ | $\sigma$ | $MSE\%$ | $W_v$ | $W_t$(min) |
| --- | --- | --- | --- | --- | --- |
| 1 | 0.9808 | 0.16225 | 0.52 | 1.3144 | — |
| 15 | 0.8139 | 0.13444 | 9.95 | 1.1905 | 2.67 |
| 30 | 0.6453 | 0.10456 | 23.5 | 1.0552 | 4.67 |
| 45 | 0.3987 | 0.09875 | 35.5 | 0.8079 | 7.37 |
| 60 | 0.3889 | 0.07361 | 44.9 | 0.7768 | 6.73 |

# 6. Discussion

In this section, we will analyze specific cases and provide explanations of the prediction made by the Crossformer in stuck pipe scenarios. Visualizations contribute to a deeper understanding of the model's performance in addressing drilling-related problems. For a black-box algorithm, providing explanation in practical applications is essential to assist drilling engineers in making informed decisions. Furthermore, providing explanations for the algorithm is valuable for subsequent remedial measures, such as identifying the reasons and types of stuck pipe incidents and performing the necessary operations to free the drill string.

**Case Study 1**: In the case of TSPP2 as utilized in the results section, the model's prediction error ($MSE$) increases with the increase of $\tau$, the values of $\boldsymbol{R}_{T:T+\Lambda+\tau;d}$ show an increasing trend with the increase in $MSE$. Although we have excluded the majority of abnormal time periods from the calculation range, the remaining values continue exert an influence on the normalized results. After normalization, this characteristic causes the phenomenon in normal drilling data where larger $\tau$ to be assigned smaller values in $Risk$, this consistency in results is in accordance with the methodology we employed. Take $\tau = 30$ for an example, the curves depicting the ground truth and predictions of all parameters as well as $Risk$ over time are plotted in Fig. 7. The values of $Risk$ gradually increase from 0.65 to 2.46 before a stuck pipe incident occurs. It is evident that the primary reconstruction error leading to an increase in $Risk$ before getting stuck are attributed to torque, WOB, and ROP.

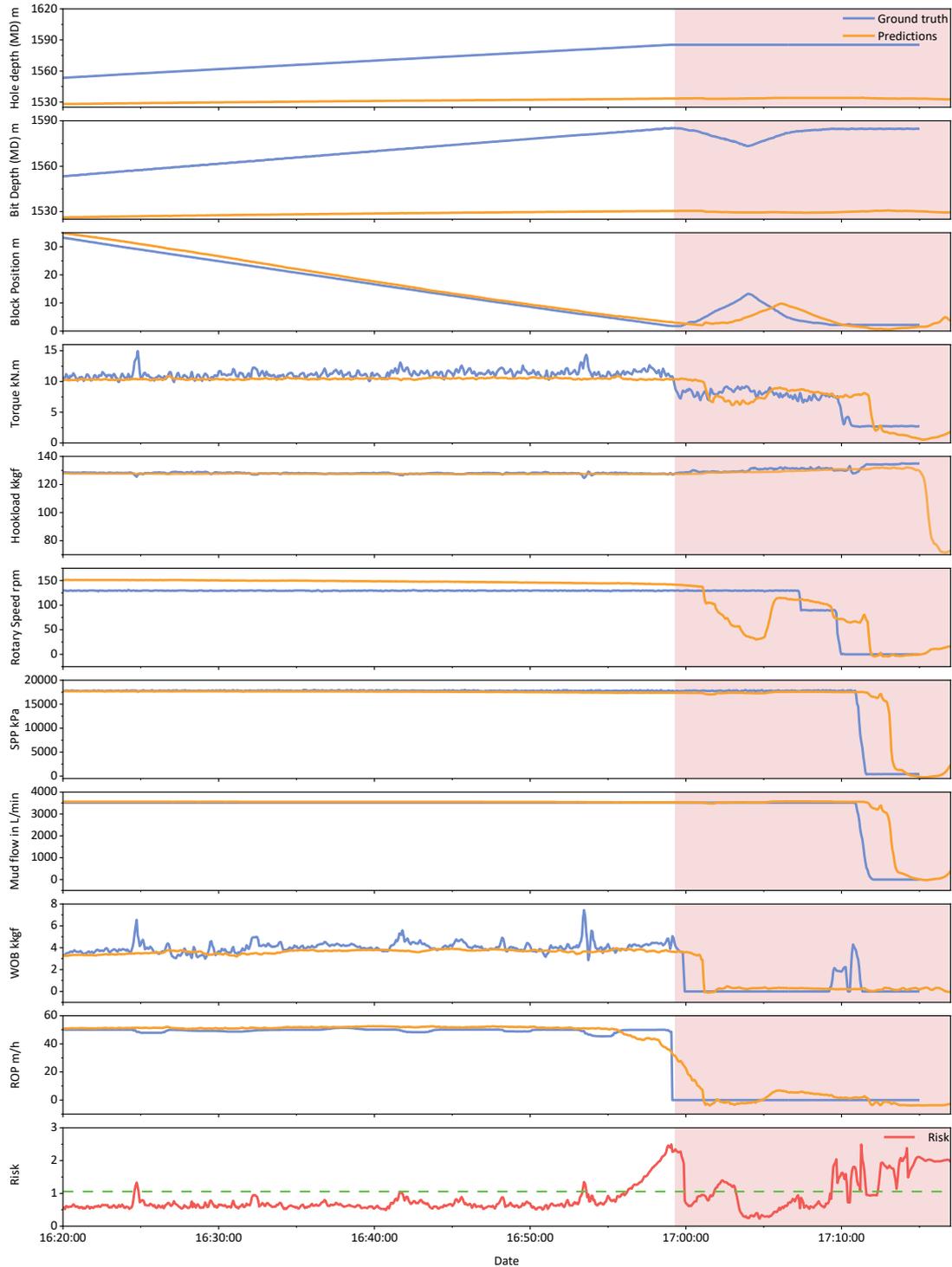

Fig. 7. For TSPP2, the model detects the early sign of a stuck pipe incident 4.67 minutes in advance when the prediction step size is set to 30, the ground truth and predictions for various parameters are indicated in blue and yellow, respectively. The warning threshold is indicated by the green dashed line.

**Case Study 2**: For the stuck pipe case from TSPP3, $\tau = 30$ is chosen. The curves showing the ground truth and predictions of various parameters changing over time are presented in Fig. 8, with the bottom panel still representing $Risk$. In this case, a stuck pipe incident occurs at 7:28, the drastic fluctuations in rotary speed and WOB actively adjusted by the driller, which also result in abnormal variations in $Risk$. The reconstruction errors of normal drilling data from 06:44 to 06:55 are selected for probability density calculation, because the rotary speed and WOB values during this period differed significantly from the values at other times, which leading a large difference in ROP according to Bingham model (Bingham, 1965), it results in a significant difference in $Risk$ between this period and other normal drilling scenarios. Our model detects that $Risk$ consistently exceed $W_v = 1.2355$ after 07:05, warning time $W_t$ is determined to be 23.60 minutes, indicating that the model successfully detects the early sign of a stuck pipe.

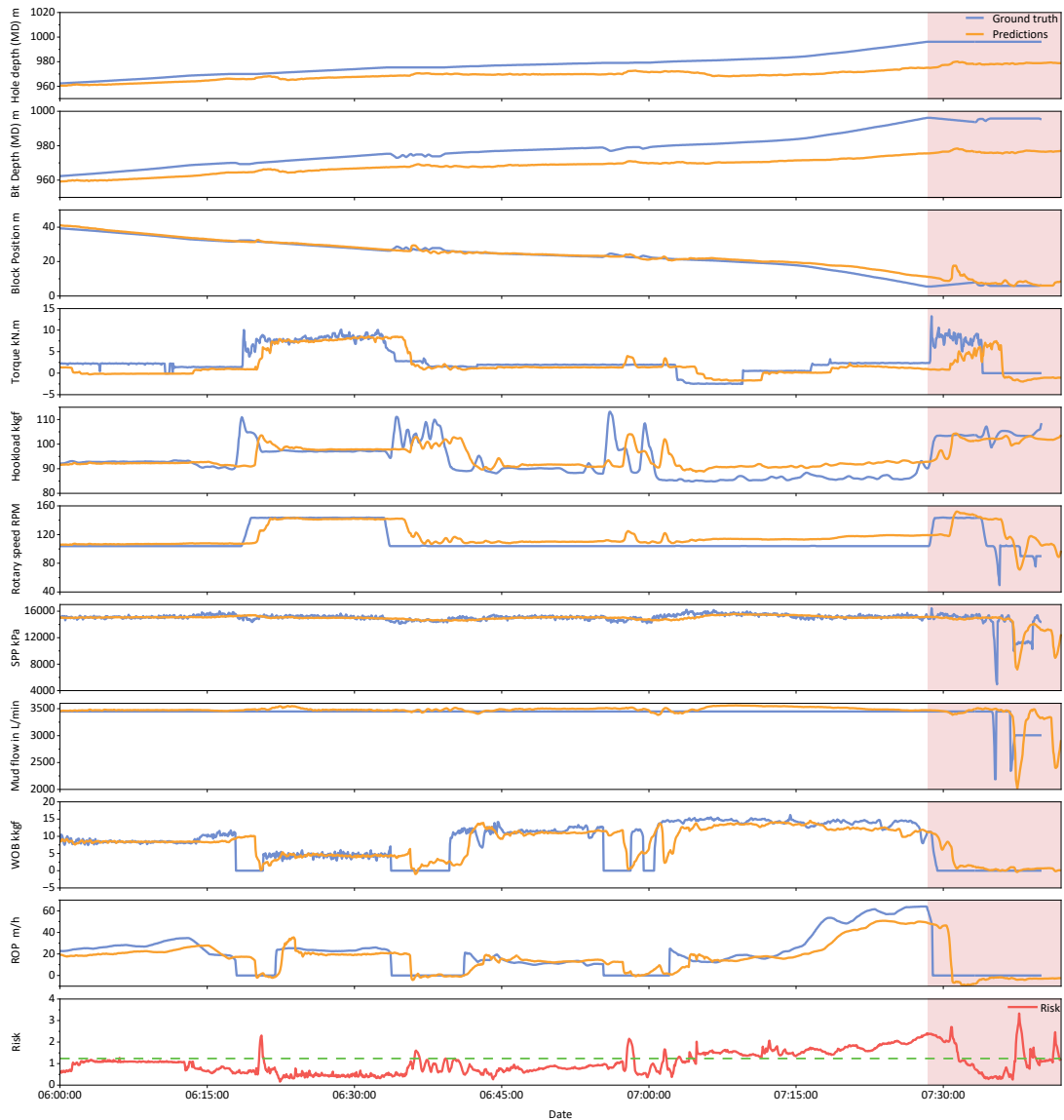

Fig. 8. The model detects the early sign of a stuck pipe incident 23.60 minutes before it occurred on TSPP3. The ground truth and predictions for various parameters are indicated in blue and yellow, respectively. The warning threshold is indicated by the green dashed line.

6.1 Explanation of the model's predictions

The predictive performance of Crossformer significantly better than Informer at the same prediction step size. Informer takes into account the sparsity of the attention layers and effectively reduces computational complexity through the combination of the self-attention distilling operation and Kullback-Leibler divergence. However, all data is valuable in drilling engineering, the introduction of Kullback-Leibler divergence discards low attention scores of data patterns, causes a decrease in the predictive ability of model.

The model did not detect the early sign of stuck pipe when $\tau = 1$. To clarify this, we need to revisit the meaning of $Risk$. $Risk$ is the relative variation rate of predictions after $\tau$ steps relative to current ground truth. We assume that the data changes continuously before being resampled, which implies that the change in data from one step to the next is not obvious. $Risk_{T+\lambda+1:T+\lambda+\tau}$ is divided into $\tau$ parts when making single-step prediction. What's even more challenging is that the model exhibits high predictive accuracy and resulting in accurate outputs, whereas our evaluation method regards reconstruction error as $Risk$. When these two factors are combined, they result in a lack of obvious difference between the $Risk$ values output by each step of sliding window in stuck pipe prediction and those observed during a normal drilling scenario when $\tau = 1$, as shown in Fig. 7. In such case, the model become overly sensitive to momentary data fluctuations, thereby failing to accurately differentiate between normal operations and early sign of stuck pipe incidents.

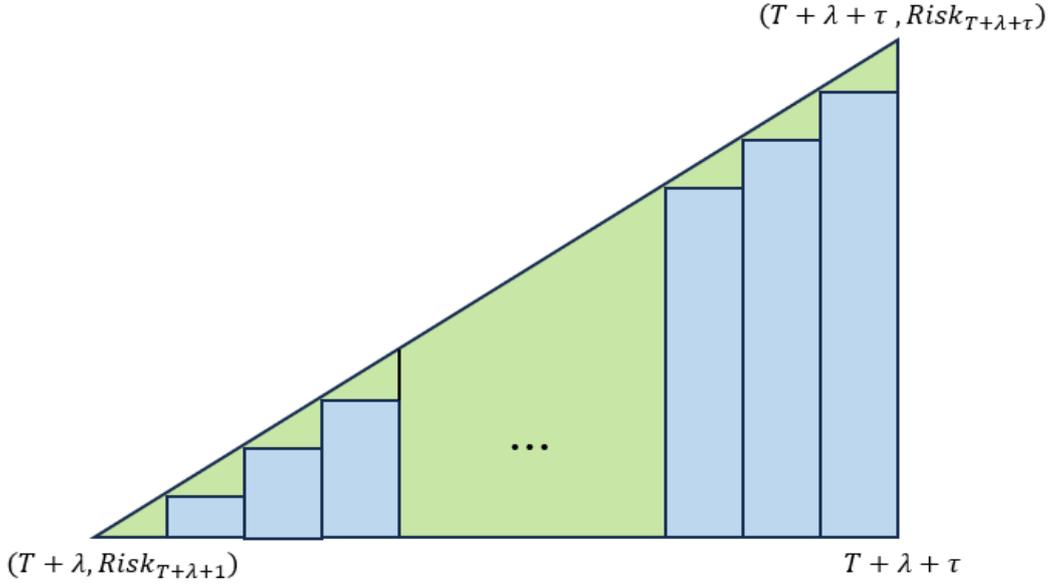

Fig. 9. The visual explanation for why early sign of stuck pipe incidents cannot be detected when $\tau = 1$. The model is unable to obtain sufficient contextual information to recognize complex patterns and trends in drilling data. Note that due to great variation in modeling error under two different prediction step sizes, the actual $Risk_{T+\lambda+30} - Risk_{T+\lambda+1} \gg 30(Risk_{T+\lambda+2} - Risk_{T+\lambda+1})$.

The model captures both cross-time dependency and cross-dimension dependency of drilling parameters. The prediction curves exhibit 2 minutes time delay phenomenon. This is due to the model predicting and displaying the current time $\tau$ steps ahead, i.e., 2 minutes, leading to a delay phenomenon in the display of the prediction curve for current changes. Furthermore, the similarity in the trend of changes between the ground truth curve and the predictions curve indirectly validates the reliability of the model. The fluctuations in the ground truth of data at the same moment among multiple dimensions demonstrate the interdependencies of drilling parameters. In case 2, the driller actively adjusted the ground truth of the drilling parameter WOB around 07:00, causing the ground truth of the parameters hookload and ROP to exhibit immediate fluctuations. However, there were no significant changes in the ground truth of the parameters rotary speed and torque. Surprisingly, the predictions of these two parameters still exhibited fluctuations. Since the ground truth of the parameters rotary speed and torque did not change, it can be concluded that the fluctuations in predictions were definitely not caused by changes in the ground truth. Therefore, the changes in predictions at this moment must be due to the model reproducing the interdependency between parameters to make predictions. This is a crucial reason for our use of Crossformer, as it helps capture and modeling these complex relationships among dimensions effectively. We hypothesize that these fluctuations may, to some extent, be explained as early indicators of a stuck pipe, potentially signing the ongoing deterioration of downhole conditions.

After comparing the two cases mentioned above, it is easily deduced that the duration of early signs for a stuck pipe has a decisive impact on the length of the warning time. In case 1, due to the excessively short duration of early signs, on-site drilling engineers might not have enough time to implement subsequent remedial measures. However, the warning time is generally considered to be ample in case 2. A sufficiently small $MSE$ forms the foundation for the model to work effectively. However, reducing the prediction step size to lower $MSE$ can lead to reduced warning time, having adequate warning time is a prerequisite for the practical application of stuck pipe prediction and is crucial for subsequent remedial measures. In order to enhance the applicability of model for on-site scenarios and provide guidance for drilling operations, a trade-off should be considered between model prediction error, parallel computing capability and warning time. Based on this thought, shorter prediction step size can be employed to reduce prediction error $\boldsymbol{E^\tau}$ thereby enhancing the accuracy and credibility of the model in case 2. Furthermore, for the same set of input data, we can even simultaneously select multiple prediction step sizes for parallel computation, yielding several mutually referenced results.

6.2 Limitations of study

Our model can only predict stuck pipe incidents that occur during the drilling process. This is because our model is trained only on normal drilling processes to learn data patterns, the model lacks predictive capability for non-drilling scenarios. There are various mechanisms of stuck, thus, there is no a single data pattern indicating all of stuck pipe incidents (Salminen et al., 2017). For drilling operations like making connections, false alarms can occur when adding drill strings, the sudden stuck pipe without early sign (TSPP4), such as junk sticking incidents, can't be predicted. Additionally, incidents of keyseat sticking that occur during the process of tripping can't be predicted.

When the early sign of a stuck pipe incident can be detected at multiple prediction step size, it is difficult to determine which prediction step size is the most appropriate. As mentioned above, a trade-off needs to be considered between prediction error, parallel computing capability and warning time to enhance the applicability of model. The warning time varies obviously among different stuck pipe incidents, we can't determine the warning time until a stuck pipe incident occurs, selecting a prediction step size after the incident has already occurred renders the prediction meaningless, which appears to be a paradox. Choosing the more appropriate prediction step size for different scenarios requires further investigation and research.

The amount of data and the number of cases used in our research are insufficient. A larger volume of training data will yield a smaller $MSE$, as shown in Table 3, because it provides a more comprehensive description of the drilling data patterns. In Case 2, the WOB values for the early sign of stuck pipe are slightly higher when compared to the sample data used to generate thresholds, which has an impact on the creation of warning threshold. However, due to limited location information between wells, we are not able to find data from neighboring wells in similar formations. In addition, conducting more case analyses also contributes to a deeper research of prediction step size selection.

# 7. Conclusions

The ability of the model to accurately modeling and reconstruct long sequences is crucial for stuck pipe prediction, existing models struggle to do so effectively. We employed improved Crossformer with sliding window technique for early signs of stuck pipe detection in drilling process. The model can accurately capture the changes in each parameter at each time step, the stuck risk at each time step is provided $\tau$ steps in advance, leading to the determination of warning threshold and warning time. The results indicate that our model can successfully detect early signs of stuck pipe for multiple prediction step sizes.

(1) The Crossformer model, which captures the dependencies among drilling parameters, exhibits superior modeling and predictive capabilities compared to other DL models.

(2) Finer data granularity and more extensive training to include a wider range of data patterns help enhance the predictive capability of the model.

(3) Compared to a single prediction step size, multi-step prediction method demonstrates superiority in predicting stuck pipe incidents.

Given the immense potential demonstrated by our methodology, for future work, we will uphold Equinor's original intent in opening the Volve field dataset to further develop the data from the Volve field, which enabling the model to receive more comprehensive training and learn a wider variety of data patterns under different drilling conditions. Since proposed algorithm offers a universal approach when calculating $Risk$, by directly summing the relative reconstruction errors of all parameters, it assumes that all drilling parameters contribute equally to the risk of stuck pipe incidents. We can provide the ability to adjust the weights for different parameter combinations to make the algorithm more sensitive to specific stuck pipe incidents, thus enabling it to identify particular types of stuck pipe incidents.

# References

Abbas, A.K., Flori, R., Almubarak, H., et al., 2019. Intelligent Prediction of Stuck Pipe Remediation Using Machine Learning Algorithms. In: SPE Annual Technical Conference and Exhibition. OnePetro, Alberta, Calgary, Canada. https://doi.org/10.2118/196229-MS.

Al-Baiyat, I., Heinze, L., 2012. Implementing Artificial Neural Networks and Support Vector Machines in Stuck Pipe Prediction. In: SPE Kuwait International Petroleum Conference and Exhibition. 10-12 December Kuwait City, Kuwait. https://doi.org/10.2118/163370-MS.

Alshaikh, A., Magana-Mora, A., Gharbi, et al., 2019. Machine Learning for Detecting Stuck Pipe Incidents: Data Analytics and Models Evaluation. In: International Petroleum Technology Conference. 26-28 March, Beijing, China. https://doi.org/10.2523/IPTC-19394-MS.

Bahdanau, D., Cho, K., Bengio, Y., 2014. Neural Machine Translation by Jointly Learning to Align and Translate. arXiv:1409.0473.

Bingham, M., 1965. A new approach to interpreting--rock drillability. The Petroleum Publishing Co.

Bourgoyne, A.T., Jr., Young, et al., 1974. A Multiple Regression Approach to Optimal Drilling and Abnormal Pressure Detection. Soc. Pet. Eng. J. 14, 371-384. https://doi.org/10.2118/4238-PA.

Child, R., Gray, S., Radford, et al., 2019. Generating Long Sequences with Sparse Transformers. arXiv: 1904.10509.

Cho, K., van Merrienboer, B., Bahdanau, et al., 2014. On the Properties of Neural Machine Translation: Encoder-Decoder. Approaches. Association for Computational Linguistics. pp. 103-111. https://doi.org/10.3115/v1/W14-4012

Equinor, 2018. Volve field data (CC BY-NC-SA 4.0). https://www.equinor.com/en/news /14jun2018-disclosing-volve-data.html.

Hochreiter, S., Schmidhuber, J., 1997. Long Short-Term Memory. Neural Comput. 9, 1735-1780. https://doi.org/10.1162/neco.1997.9.8.1735.

Inoue, T., Nakagawa, Y., Wada, R., et al., 2022. Early Stuck Detection Using Supervised and Unsupervised Machine Learning Approaches. In: Offshore Technology Conference Asia. 22-25 March, Virtual and Kuala Lumpur, Malaysia. https://doi.org/10.4043/31376-MS.

Jahanbakhshi, R., Keshavarzi, R., Aliyari Shoorehdeli, M., et al., 2012. Intelligent Prediction of Differential Pipe Sticking by Support Vector Machine Compared With Conventional Artificial Neural Networks: An Example of Iranian Offshore Oil Fields. SPE Drill. Complet. 27, 586-595. https://doi.org/10.2118/163062-PA.

Li, S., Jin, X., Xuan, Y., et al., 2020. Enhancing the Locality and Breaking the Memory Bottleneck of Transformer on Time Series Forecasting. In: Proceedings of the 33rd International Conference on Neural Information Processing Systems. 471, 52

43-5253.

Magana-Mora, A., Gharbi, S., Alshaikh, A., et al., 2019. AccuPipePred: A Framework for the Accurate and Early Detection of Stuck Pipe for Real-Time Drilling Operations. In: SPE Middle East Oil and Gas Show and Conference. 18-21 March, Manama, Bahrain. https://doi.org/10.2118/194980-MS.

Mopuri, K.R., Bilen, H., Tsuchihashi, N., et al., 2022. Early sign detection for the stuck pipe scenarios using unsupervised deep learning. J. Pet. Sci. Eng. 208, 109489. https://doi.org/10.1016/j.petrol.2021.109489.

Muqeem, M.A., Weekse, A.E., Al-Hajji, A.A., 2012. In: SPE Saudi Arabia Section Technical Symposium and Exhibition. 8-11 April, Al-Khobar, Saudi Arabia. https://doi.org/10.2118/160845-MS.

Tsuchihashi, N., Wada, R., Ozaki, M., et al., 2021. Early stuck pipe sign detection with depth-domain 3D convolutional neural network using actual drilling data. SPE Journal, 26(02), 551-562. https://doi.org/10.2118/204462-PA

Salminen, K., Cheatham, C., Smith, M., et al., 2017. Stuck-Pipe Prediction by Use of Automated Real-Time Modeling and Data Analysis. SPE Drill. Complet. 32, 184-193. https://doi.org/10.2118/178888-PA.

Siqueira, V.S.D.M., Cuadros, M.A.S., 2024. Expert system for early sign stuck pipe detection: Feature engineering and fuzzy logic approach. Engineering Applications of Artificial Intelligence, 127 (2024): 107229. https://doi.org/10.1016/j.engappai.2023.107229

Song, H., Rajan, D., Thiagarajan, J.J., et al., 2017. Attend and Diagnose: Clinical Time Series Analysis using Attention Models. In: Proceedings of the AAAI Conference on Artificial Intelligence. 32, 1. https://doi.org/10.1609/aaai.v32i1.11635.

Sutskever, I., Vinyals, O., Le, Q., 2014. Sequence to Sequence Learning with Neural Networks. In: Advances in neural information processing systems.

Tunkiel, A., Wiktorski, T., Sui, D., 2020. Drilling Dataset Exploration, Processing and Interpretation Using Volve Field Data. In: International Conference on Offshore Mechanics and Arctic Engineering. June 28-July 3, Fort Lauderdale, FL, USA. https://doi.org/10.1115/OMAE2020-18151

Tunkiel, A.T., Sui, D., Wiktorski, T., 2021. Training-while-drilling approach to inclination prediction in directional drilling utilizing recurrent neural networks. J. Pet. Sci. Eng. 196, 108128. https://doi.org/10.1016/j.petrol.2020.108128.

Unrau, S., Torrione, P., Hibbard, M., et al. 2017. Machine Learning Algorithms Applied to Detection of Well Control Events. In: SPE Kingdom of Saudi Arabia Annual Technical Symposium and Exhibition. 24-27 April, Dammam, Saudi Arabia. https://doi.org/10.2118/188104-MS.

Vaswani, A., Shazeer, N., Parmar, N., et al., 2017. Attention Is All You Need. In: Advances in Neural Information Processing Systems.

Wu, H., Xu, J., Wang, J., et al., 2022. Autoformer: Decomposition Transformers with Auto-Correlation for Long-Term Series Forecasting. In: Advances in Neural Information Processing Systems. pp. 22419-22430.

Xu, J., Wu, H., Wang, J., et al., 2021. Anomaly transformer: Time series anomaly detection with association discrepancy. arXiv:2110.02642.

Yang, C.-H.H., Tsai, Y.-Y., Chen, P.-Y., 2021. Voice2Series: Reprogramming Acoustic Models for Time Series Classification, In: International Conference on Machine Learning. pp. 11808-11819.

Zerveas, G., Jayaraman, S., Patel, D., 2021. A Transformer-based Framework for Multivariate Time Series Representation Learning. In: Association for Computing Machinery. pp. 2114-2124. https://doi.org/10.1145/3447548.3467401.

Zhang, Y., Yan, J., 2023. Crossformer: Transformer Utilizing Cross-dimension Dependency for Multivariate Time Series Forecasting. In: International Conference on Learning Representations.

Zhou, H., Li, J., Zhang, S., et al., 2023. Expanding the prediction capacity in long sequence time-series forecasting. Artif. Intell. 318, 103886. https://doi.org/10.1016/j.artint.2023.103886.

Zhou, T., Ma, Z., Wen, Q., et al., 2022. FEDformer: Frequency Enhanced Decomposed Transformer for Long-term Series Forecasting. In: International Conference on Machine Learning. pp. 27268-27286.